\documentclass[twocolumn,english]{IEEEtran}
\usepackage[T1]{fontenc}
\usepackage[latin9]{inputenc}
\usepackage{babel}
\usepackage{cite}
\usepackage{textcomp}
\usepackage{amsmath}
\usepackage{amssymb}
\usepackage{graphicx}
\usepackage{xcolor}
\usepackage[unicode=true, bookmarks=true, bookmarksnumbered=true, bookmarksopen=true, bookmarksopenlevel=1, breaklinks=false, pdfborder={0 0 1}, backref=false, colorlinks=false]{hyperref}
\hypersetup{pdftitle={Your Title}, pdfauthor={Your Name}, pdfborderstyle=, pdfpagelayout=OneColumn, pdfnewwindow=true, pdfstartview=XYZ, plainpages=false}

\makeatletter
\usepackage{babel}
\usepackage{units}
\usepackage{graphicx}
\usepackage{subcaption}

\ifCLASSOPTIONcompsoc
\else
\fi

\global\long\def\ie{\textit{i.e.,} }

\makeatother

\begin{document}
\title{On the impact of the antenna radiation patterns \\
 in passive radio sensing
\thanks{This work is funded by the EU. Grant Agreement No: 101099491}}
\author{Federica Fieramosca, Vittorio Rampa, \textit{Senior Member, IEEE},
Stefano Savazzi, \textit{Member, IEEE}, \\
 and Michele D'Amico, \textit{Senior Member, IEEE} \thanks{F. Fieramosca and M. D'Amico are with Politecnico di Milano, DEIB
department, e-mail: \{federica.fieramosca,michele.damico\}@polimi.it.} \thanks{V. Rampa and S. Savazzi are with Consiglio Nazionale delle Ricerche
(CNR), IEIIT institute, e-mail: \{vittorio.rampa,stefano.savazzi\}@ieiit.cnr.it.} }
\maketitle
\begin{abstract}
Electromagnetic (EM) body models based on the scalar diffraction theory
allow to predict the impact of subject motions on the radio propagation
channel without requiring a time-consuming full-wave approach. On
the other hand, they are less effective in complex environments characterized
by significant multipath effects. Recently, emerging radio sensing
applications have proposed the adoption of smart antennas with non-isotropic
radiation characteristics to improve coverage. 
This letter investigates the impact of antenna radiation patterns
in passive radio sensing applications. Adaptations of diffraction-based
EM models are proposed to account for antenna non-uniform angular
filtering. Next, we quantify experimentally the impact of diffraction
and multipath disturbance components on radio sensing accuracy in
environments with smart antennas.


\thispagestyle{empty} 
\end{abstract}


\section{Introduction}

\IEEEPARstart{P}{assive} or device-free radio sensing is an opportunistic
technique that employs stray ambient signals from Radio Frequency
(RF) devices to detect, locate, and track people that do not carry
any electronic device \cite{youssef-2007,savazzi-2016}. 
The effect of the presence of body obstacles on the received RF signals
is a well-know topic in the wireless communications community \cite{brittain-1994,krupka-1968,king-1977}.
However, only recently, radio sensing techniques have been proposed
to provide sensing capabilities, while performing radio communication
according to the \emph{Communication while Sensing} paradigm \cite{savazzi-2016}.

Quantitative evaluation \cite{ghaddar-2004,ghaddar-2007,koutatis-2010,wang-2015,rampa-2017,rampa-2022a}
of perturbations due to the presence or movements of people (\ie
the targets) has paved the way to the exploitation of electromagnetic
(EM) models for passive radio sensing. In fact, the body-induced perturbations
that impair the radio channel, can be acquired, measured, and processed
using model-based methods to estimate location \cite{shit-2019},
and tracking target info \cite{wang-2015,kalt-2021}, or to assess
location accuracy during network pre-deployment \cite{kianoush-2016}.

However, a general EM model for the prediction of body-induced effects
on propagation 
is still under scrutiny \cite{hamilton-2014}, or too complex to be
of practical use for real-time sensing scenarios \cite{ghaddar-2007,eleryan-2011}.
Simpler human-body shadowing models have been recently proposed for
\color{black}Device-Free Localization (DFL) \color{black} based
on scalar diffraction theory \cite{rampa-2017,rampa-2022a}. 

Other semi-empirical models \cite{nannuru-2012,guo-2014,access-2021}
have been also proposed for DFL applications \cite{wilson-2010,mohamed-2017,wang-2012,sukor-2020}.
However, these models require lengthy calibration pre-processing steps
and will not be considered here (see \cite{savazzi-2016,shit-2019}
for references).

\begin{figure}
\begin{centering}
\includegraphics[scale=0.35]{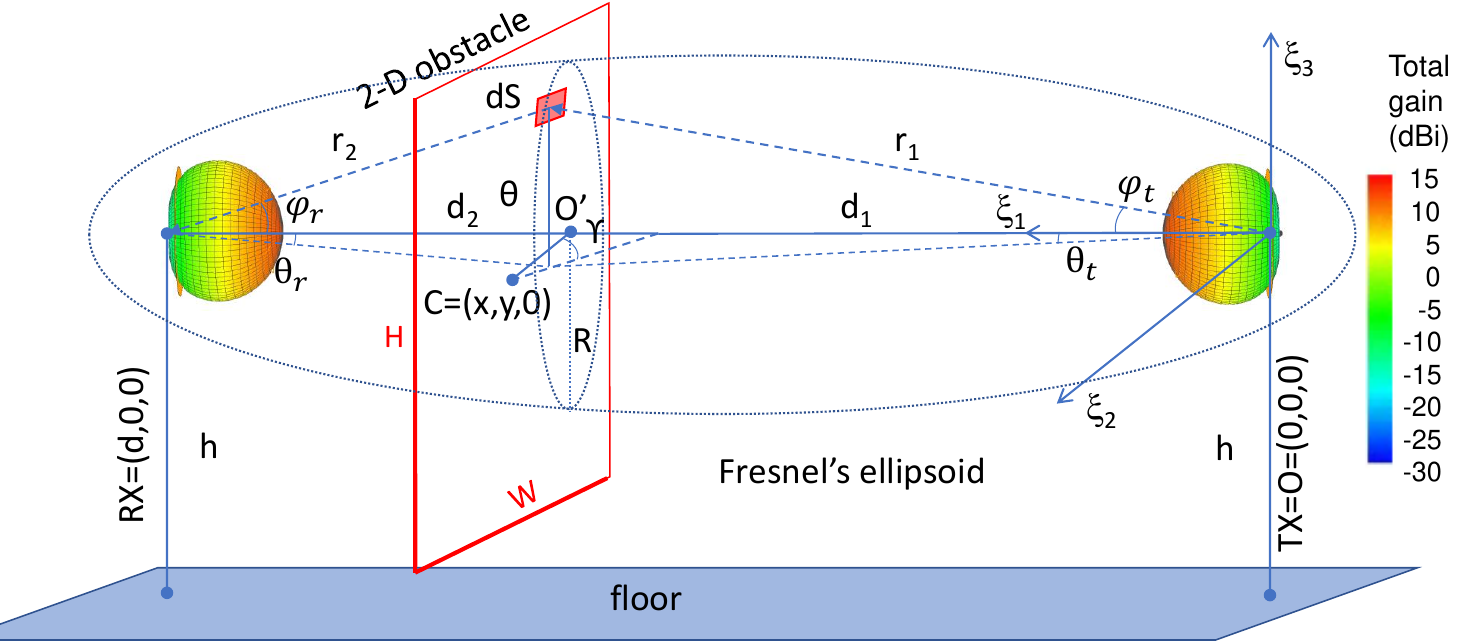} 
\par\end{centering}
\caption{\label{fig:radiation_patterns_link}\color{black}EM model geometry:
$2$-D obstacle and antennas.}
\end{figure}
\color{black}

\section{Paper contributions}

Considering the interest in novel Wireless Local Area Network (WLAN)
sensing systems \cite{rampa-2022b,halperin-2011,xie-2018,chauhan-2021}
and tools \cite{atif-2020}, with devices leveraging antennas with
non-uniform \cite{zhang-2018,shukri-2019,garcia-2020}, and/or re-configurable
\cite{santoboni-2022} radiation characteristics, it is deemed necessary
to develop effective EM models \cite{access-2021,kalt-2021} that
meet these emerging needs. Most of the previous tools were based on
diffraction methods \cite{wang-2015,rampa-2022a,eleryan-2011}
and targeted devices equipped with omnidirectional antennas, with
the exception of \cite{maccartney-2016} that focused on human blockage
at $73$~GHz with the body represented as a semi-infinite rectangular
shape and the paraxial approximation \cite{rampa-2017} being used.

The key ideas discussed in this letter are: \emph{i}) the proposal
of a simple human-body shadowing model, that includes also the antenna
directivity characteristics; \emph{ii}) the application of the proposed
model in passive radio sensing and validation of its predictive potential;
and \emph{iii}) the evaluation of the impact of antenna radiation
patterns by exploiting real on-field measurements in an indoor reflective
environment. 
The paper is organized as follows: Sect. \ref{sec:Diffraction-models}
presents a EM body model that includes the directional radiation pattern
hypothesis while Sect. \ref{sec:Physical-statistical-modeling} analyzes
the body-induced effects in scenarios with mixed antenna systems (\ie
both directional and omnidirectional). Sect. \ref{sec:Model-validation}
validates the proposed body model in real field scenarios. Finally,
Sect. \ref{sec:Conclusions} draws some conclusions and proposes additional
investigations.

\section{EM body models \label{sec:Diffraction-models}}

In this work, the statistical body model proposed in \cite{rampa-2017}
for isotropic antennas is extended to take into account directional
antennas with an assigned radiation pattern. We consider a single
body, but the extension to multi-body scenarios can be inferred
according to \cite{rampa-2022a,rampa-2019}. \color{black} We
also assume that the body is in the Fraunhofer's regions of the
antennas of the transmitter (TX) and receiver (RX): the regions start
$\approx 25$~cm away from the directional and $\approx 15$~cm from the omnidirectional antennas of the experimental setup shown in Sect. \ref{sec:Physical-statistical-modeling}. \color{black} 

As shown in Fig.~\ref{fig:radiation_patterns_link}, the 3-D shape
of the human body is modeled as a $2$-D rectangular absorbing sheet
$S$ \cite{rampa-2017} of height $H$ and traversal size $W$, and has nominal position coordinates $(x,y)$, w.r.t. the TX position, namely the projection of its barycenter on the horizontal
plane. The scalar diffraction theory \cite{mokhtari-1999,durgin-2009}
quantifies the impact of this obstruction. First, a distribution of
Huygens' sources of elementary area $dS$ is assumed to be located
on the absorbing sheet. Then, the electric field $E=E_{0}-\int_{S}dE$
at the RX is obtained by subtracting the contribution of
the Huygens' sources $\int_{S}dE$ from the electric field $E_{0}$
of the free-space scenario (\ie with no target in the link area).

Using the received electric field $E_{0}$ under free-space condition
as reference, for both isotropic antennas, we get\cite{rampa-2017}:
\begin{equation}
\frac{E}{E_{0}}=1-j\,\frac{d}{\lambda}\,\int_{S}\frac{1}{r_{1}\,r_{2}}\,\exp\left\{ -j\frac{2\pi}{\lambda}\left(r_{1}+r_{2}-d\right)\right\} d\xi_{2}\,d\xi_{3},\label{eq:dE_full_compact-1-1-1}
\end{equation}
where $d$ is the link length, \color{black}$\lambda=c/f$ is the wavelength, while $f$ is the frequency and $c$ is the speed of light\color{black}. Notice
that each elementary source $dS=d\xi_{2}\,d\xi_{3}$ has distance
$r_{1}$ and $r_{2}$ from the TX and RX, respectively. 

The received power $P$ is defined at the generic frequency $f$,
omitted here for clarity, as: 
\begin{equation}
P=\left\{ \begin{array}{ll}
P_{0}+w_{0} & \;\textrm{free-space only}\\
P_{0}-A_{\textrm{S}}(x,y)+w_{T} & \;\textrm{with target }S,
\end{array}\right.\label{eq:received_power}
\end{equation}
where $A_{\textrm{S}}(x,y)=-10\,\log_{10}\left|E\,/\,E_{0}\right|^{2}$
is the extra-attenuation due to the presence of $S$ at coordinates
$(x,y)$. The free-space power $P_{0}$ is a constant that depends
only on the link geometry and on the propagation coefficients: it
is assumed to be known, or measured. The log-normal multipath fading
and the other disturbances are modeled as the Gaussian noise terms
$w_{0}\sim\mathcal{N\mathrm{\left(0,\mathit{\sigma_{\textrm{0}}^{2}}\right)}}$,
with variance $\mathrm{\mathit{\sigma_{\textrm{0}}^{2}}}$, and $w_{T}\sim\mathcal{N\mathrm{\left(\mathit{\mu_{T},\sigma_{T}^{2}}\right)}}$,
with mean $\mu_{T}=\Delta h_{T}$ and variance $\mathrm{\mathit{\sigma_{T}^{2}}}=\mathit{\sigma_{\textrm{0}}^{2}}+\Delta\mathit{\sigma_{T}^{2}}$,
respectively. $\Delta h_{T}$ and $\Delta\sigma_{T}^{2}\geq0$ are
the residual stochastic fading terms that depend on the specific scenario
as in \cite{rampa-2017}.

For a generic non-isotropic antenna, equation (\ref{eq:dE_full_compact-1-1-1})
must be modified to take into account the antenna radiation pattern
$G\left(\theta,\varphi\right)=G_{0}\,f\left(\theta,\varphi\right)$,
where $G_{0}$ is the gain and $f\left(\theta,\varphi\right)$ is
the normalized radiation pattern, while $\theta$ and
$\varphi$ are the polar coordinates, usually referred to the antenna
phase center. First, we  consider an isotropic RX antenna
and a directional TX one that is pointed in the Line Of Sight (LOS)
direction, with normalized radiation pattern $f_{t}\left(\theta_{t},\varphi_{t}\right)$
and \color{black} polar coordinates $\theta_{t}=\theta_{t}\left(r_{1},r_{2}\right)$
and $\varphi_{t}=\varphi_{t}\left(r_{1},r_{2}\right)$ w.r.t. the
TX antenna phase center. \color{black} The field ratio $E/E_{0}$
in (\ref{eq:dE_full_compact-1-1-1}) becomes: 
\begin{equation}
\begin{aligned}\frac{E}{E_{0}}={} & 1-j\,\frac{d}{\lambda}\,\int_{S}\frac{1}{r_{1}\,r_{2}}\,\sqrt{f_{t}\left(\theta_{t},\varphi_{t}\right)}\,\cdot\\
{} & \cdot\,\exp\left\{ -j\frac{2\pi}{\lambda}\left(r_{1}+r_{2}-d\right)\right\} d\xi_{2}\,d\xi_{3}.
\end{aligned}
\label{eq:E_E0_approx-single}
\end{equation}

If the receiving antenna is also directional and pointed toward the
transmitter in the LOS direction, the received signal can be calculated,
with good approximation, by weighting the contributions from the elementary
Huygens' sources with the square root of the receiving antenna radiation
pattern. If $V$ and $V_{0}$ are the complex voltages at the RX antenna
connector in the actual scenario and in free-space, respectively,
we get: 
\begin{equation}
\begin{aligned}\frac{V}{V_{0}}={} & 1-j\,\frac{d}{\lambda}\,\int_{S}\frac{1}{r_{1}\,r_{2}}\,\sqrt{f_{t}\left(\theta_{t},\varphi_{t}\right)\,f_{r}\left(\theta_{r},\varphi_{r}\right)}\,\cdot\\
{} & \cdot\,\exp\left\{ -j\frac{2\pi}{\lambda}\left(r_{1}+r_{2}-d\right)\right\} d\xi_{2}\,d\xi_{3},
\end{aligned}
\label{eq:V_V0_approx-single}
\end{equation}
where $\theta_{r}=\theta_{r}\left(r_{1},r_{2}\right)$ and $\varphi_{r}=\varphi_{r}\left(r_{1},r_{2}\right)$
\color{black} are the polar coordinates w.r.t. the receiving antenna
phase center. \color{black} Eq. (\ref{eq:V_V0_approx-single}) is
derived from (\ref{eq:E_E0_approx-single}) by noting that $V$ and
$V_{0}$ are linearly dependent on $E$ and $E_{0}$, respectively,
through the effective antenna length. In this link configuration,
the extra-attenuation for target in $(x,y)$ is now given by $A_{\textrm{S}}(x,y)=-10\,\log_{10}\left|V\,/\,V_{0}\right|^{2}$.

\begin{table}[tp]
\protect\caption{\label{parameters} SA settings and directional antenna specs.}
\vspace{0cm}

\begin{centering}
\begin{tabular}{|c|c||c|c|}
\hline 
\multicolumn{2}{|c||}{\textbf{Spectrum analyzer settings}} & \multicolumn{2}{c|}{\textbf{Antenna specs}}\tabularnewline
\hline 
\textbf{Start/Stop Frequency}  & $2.4/2.5$~GHz  & \textbf{HPBW ($\theta$)}  & H: $60{^{\circ}}$ \tabularnewline
\hline 
\textbf{Frequency spacing}  & $1.25$ MHz  & \textbf{HPBW ($\varphi$)}  & V: $76{^{\circ}}$ \tabularnewline
\hline 
\textbf{Resolution BW}  & $100$ kHz  & \textbf{Polarization}  & Vertical \tabularnewline
\hline 
\textbf{TX output power}  & $0$ dBm  & \textbf{Antenna gain}  & $9$ dBi\tabularnewline
\hline 
\end{tabular}
\par\end{centering}
\medskip{}
 \vspace{-0.6cm}
 
\end{table}

\section{Body-induced effects with mixed antennas \label{sec:Physical-statistical-modeling}}

\begin{figure}
\begin{centering}
\includegraphics[scale=0.3]{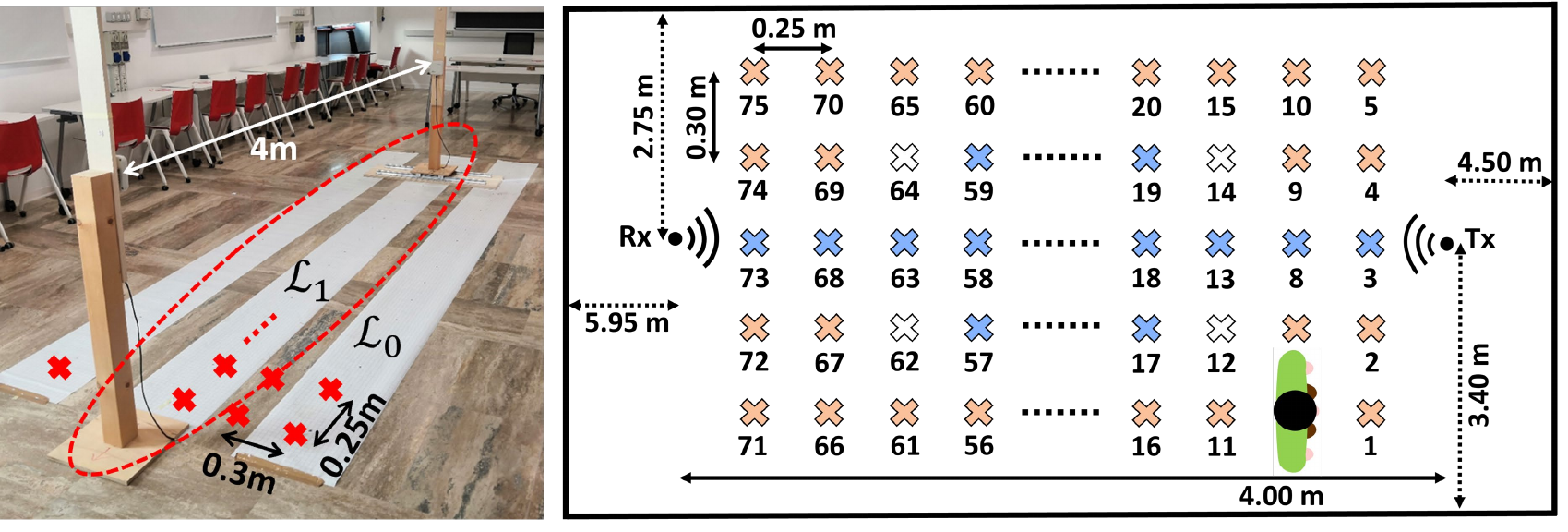} 
\par\end{centering}
\caption{\label{fig:layout}$75$ marked positions \color{black} (crosses)
on a $15\times5$ grid with spacing $0.25$~m along and $0.30$~m
across the link. Target is located at position $6$ (drawing not to
scale). Corresponding measurement scenario is on the left.}
\vspace{-0.2cm}
\end{figure}

The measurement sessions took place in a hall with size $6.15$~m
$\times$ $14.45$~m and floor-ceiling height equal to $3.35$~m.
As shown in Fig. \ref{fig:layout}, TX and RX nodes are spaced $d=4.00$~m
apart, while the LOS is horizontally placed at $h=0.99$~m from the
floor. Most surfaces are highly reflective, which cause poor DFL performances
with omnidirectional antennas \cite{rampa-2017}. The goal is to verify
the predictive capacity of the model in such complex conditions. The
received power $P$ is measured using a real-time Spectrum Analyzer
(SA) \cite{tek} with a built-in tracking generator. The SA tracks
$N_{f}=401$ frequency points equally spaced with $\Delta f=1.25$~MHz
and settings as in Tab.~\ref{parameters}. 

In what follows, three scenarios are analyzed featuring: \emph{i)}
the \textit{omni-omni}, where both TX and RX antennas are omnidirectional;
\emph{ii)} the \textit{omni-dir}, where only the TX is equipped with
a directional antenna; and \emph{iii)} the \textit{dir-dir}, where
both antennas are directional. 
Directional antennas operate at frequency band $2.4-2.5$~GHz: other
specs \cite{antenna_spec} are summarized in Tab.~\ref{parameters}.
Omnidirectional antennas are vertical monopoles with $2$~dBi gain.
To compare the measurements against the model predictions, we modeled
the body as an absorbing rectangular $2$-D sheet with height $H=2.0$~m
and traversal size $W=0.55$~m (see Fig.~\ref{fig:radiation_patterns_link}).
The maximum transversal size (\ie minor axis) of the first Fresnel's
ellipsoid is about $0.70$~m, while the beam width (at $-3$~dB)
of each directional antenna is about $2$~m at the same point ($d_{1}$=$d_{2}$=$d/2$).

The free-space received power $P_{0}(f_{k})$ is obtained for each
frequency of the set $\left\{ f_{k}\right\} _{k=1}^{N_{f}}$. 
The received power $P(f_{k},\ell)$ is then measured with the target
located in each of the $\ell=1,...,75$ marked positions of the grid
points of Fig.~\ref{fig:layout}. \color{black} Each position $\ell$
has coordinates $(x_{\ell},y_{\ell})$ with spacing $0.25$~m along
and $0.3$~m across the LOS. \color{black} The measured attenuation,
due to the target in the $\ell$-th position, is evaluated \color{black} for each
$f_{k}$ \color{black} as $A_{\textrm{S},k}^{(m)}(\ell)=-10\,\log_{10}\left[P(f_{k},\ell)\,/\,P_{0}(f_{k})\right]$
and then averaged to obtain the mean attenuation $A_{\mathrm{S}}^{(m)}(\ell)=1/N_{f}\,\sum_{k=1}^{N_{f}}A_{\mathrm{S},k}^{(m)}(\ell)$.

\begin{figure}[!t]
\centering

\subfloat{\includegraphics[angle=-90,width=1\columnwidth]{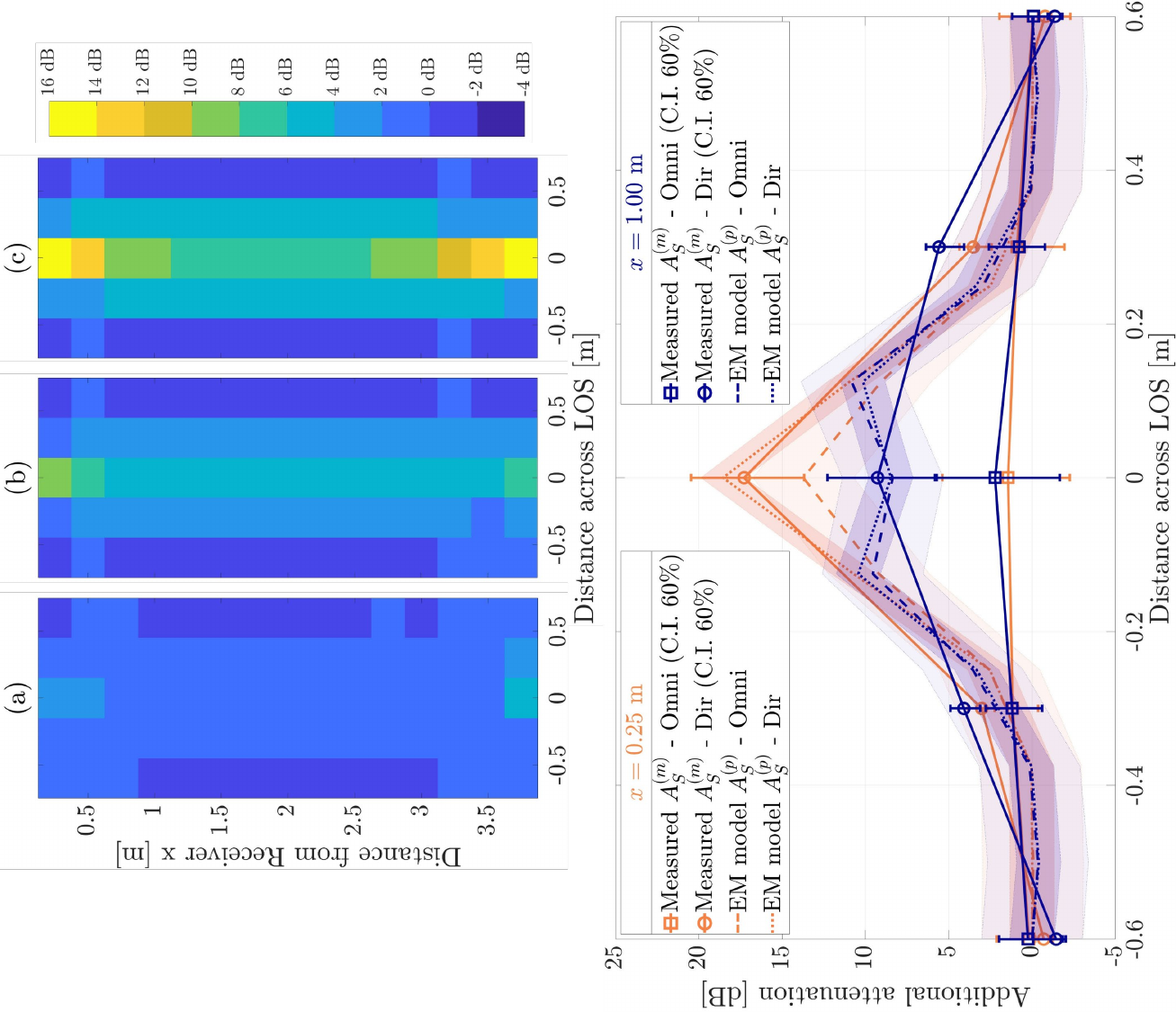}}

\caption{Top: maps of the measured attenuation (in dB) for each of the $75$
points of the a) \textit{omni-omni}, b) \textit{omni-dir}, and c)
\textit{dir-dir} scenarios. Bottom: measured $A_{\mathrm{S}}^{(m)}$
(dashed) vs. predicted $A_{\mathrm{S}}^{(p)}$ (solid) average attenuations
for a target traversing orthogonal to the LOS at $0.25$~m (orange)
and $1$~m (violet) away from the TX. Model (\ref{eq:dE_full_compact-1-1-1})
with square markers and (\ref{eq:V_V0_approx-single}) with cross
markers.}
\label{fig:maps} 
\vspace{-0.2cm}
\end{figure}

The color-coded maps in Figs.~\ref{fig:maps}.a, ~\ref{fig:maps}.b, and ~\ref{fig:maps}.c show the attenuation values for each subject
position. 
For the \textit{omni-omni} case of Fig.~\ref{fig:maps}.a, the maximum
value of attenuation is $\approx4$~dB. The body effect is thus
negligible, except for positions very close to the antennas,  
\color{black} due to a substantial amount of  energy that reaches the RX antenna via multipath even if the first Fresnel's ellipsoid is blocked\color{black}.
On the contrary, in the \textit{dir-dir} scenario of Fig.~\ref{fig:maps}.c,
the maximum attenuation reaches $\approx 16$~dB, and the body presence near the LOS is clearly discernible. In fact, by using
well-pointed directional antennas, the multipath impact is strongly
reduced thanks to the angular filtering properties of the radiation
patterns $f\left(\theta,\varphi\right)$. This scenario is thus closer
to the ideal free-space environment with no disturbances. The \textit{omni-dir}
scenario of Fig.~\ref{fig:maps}.b shows an intermediate behavior
for some noticeable effects caused by multipath disturbances not filtered by the RX antenna. The maximum attenuation reaches
$\approx 10$~dB near the TX. 

Measurements and predictions for the \emph{omni} (\ref{eq:dE_full_compact-1-1-1})
and \emph{dir} (\ref{eq:V_V0_approx-single}) setups are compared in Fig.~\ref{fig:maps} (bottom). The predictions are obtained
by averaging $A_{\mathrm{S}}^{(p)}(\ell)=1/N_{p}\,\sum_{k=1}^{N_{p}}A_{\mathrm{S}}(x_{\ell}+\triangle x_{k},y_{\ell}+\triangle y_{k})$
over the attenuations $A_{\mathrm{S}}(\cdot,\cdot)$ corresponding
to $N_{p}$ small body movements around the marked positions $\ell$. The goal is to let the models account for body position uncertainties
as well as small, involuntary movements typically observed in human sensing \cite{rampa-2017,rampa-2022a}. We set $\triangle x_{k},\triangle y_{k}\sim\mathcal{U}_{-\frac{\triangle}{2},\frac{\triangle}{2}}$
as uniformly distributed in the interval $\triangle=\,6$~cm, and
$N_{p}=150$. The measured $A_{\mathrm{S}}^{(m)}(\ell)$ (dashed lines)
and the predicted $A_{\mathrm{S}}^{(p)}(\ell)$ (solid lines) average
attenuations are compared w.r.t. $5$ marked positions along two orthogonal
cuts taken $0.25$~m (orange lines) and $1$~m (violet lines) from the
TX antenna, respectively, with marks $\ell=1\div5$ and $\ell=16\div20$
(Fig.~\ref{fig:layout}). The vertical bars include $60\%$ of the
measured values that cover the antenna operating band of $2.4-2.5$~GHz
($N_{f}=81$). Accordingly, EM predictions are obtained for $f_{k}$ in the same $2.4-2.5$~GHz band but use the field ratio (\ref{eq:dE_full_compact-1-1-1})
for omnidirectional antennas (square markers) and (\ref{eq:V_V0_approx-single}) for
directional ones (cross markers). Shaded
areas include $60\%$ of the attenuation samples used to obtain the average terms $A_{\mathrm{S}}^{(p)}(\ell)$. Overall, the measurements
reveal large fluctuations of the attenuations when the target is near
the LOS path, while the \textit{dir-dir} setup is close (on average)
to the directional antenna predictions. In general, there is a negligible
difference between \emph{omni} and \emph{dir} models when
the target is far from the TX ($x_{\ell}>1$~m) since the extra-attenuation
is mainly due to the blockage of the first Fresnel's ellipsoid. Instead,
a more marked difference is observed when the target moves close to
the TX ($x_{\ell}=0.25$~m) since the antenna beamwidth is now comparable
with the Fresnel's area. The \emph{omni} model over-estimates the attenuation
obtained from the \emph{omni-omni} setup due to the presence of multipath, as explained before. 

\section{Body detection and model validation \label{sec:Model-validation}}

We discuss here the problem of passive body localization in the environment
previously analyzed. The detection problem focuses on the choice between
the hypotheses $\mathrm{F}_{0}$ and $\mathrm{F}_{1}$ that correspond
to the target outside or inside the Fresnel's ellipsoid of the link,
respectively. According to Fig.~\ref{fig:layout}, we split the $75$
inspected positions in two groups: namely, the $|\mathcal{L}_{1}|=L_{1}=25$
positions ($\ell\in\mathcal{L}_{1}$, blue crosses) that fall inside
the Fresnel's ellipsoid, and the $|\mathcal{L}_{0}|=L_{0}=38$ positions
($\ell\in\mathcal{L}_{0}$, red crosses) that fall outside. At time
$t$, the decision whether the target is inside or outside the Fresnel's
ellipsoid is based on the extra-attenuation $A_{\textrm{S}}=P_{0}-P(t)$
that is observed w.r.t. the free-space power $P_{0}$ (in dBm). Omitting
time $t$ for clarity, the Log-Likelihood Ratio (LLR): 
\begin{equation}
\mathrm{\Gamma}(A_{\textrm{S}})=\mathrm{log}\left[\frac{\Pr\left(A_{\textrm{S}}\,|\mathrm{F}_{1}\right)}{\Pr\left(A_{\textrm{S}}\,|\mathrm{F}_{0}\right)}\right]\label{eq:llr}
\end{equation}
is used to discriminate (via thresholding on $\mathrm{\Gamma}$) between
both hypotheses. Probabilities $\Pr\left(A_{\textrm{S}}\,|\mathrm{F}_{0}\right)\sim\mathcal{N}(\mu_{\mathrm{F}_{0}},\sigma_{\mathrm{F}_{0}}^{2})$
and $\Pr\left(A_{\textrm{S}}\,|\mathrm{F}_{1}\right)\sim\mathcal{N}(\mu_{\mathrm{F}_{1}},\sigma_{\mathrm{F}_{1}}^{2})$
are log-normal distributed. The parameters $\mu_{\mathrm{F}_{0}}$
and $\mu_{\mathrm{F}_{1}}$ model the average attenuations terms,
while $\sigma_{\mathrm{F}_{0}}=\sigma_{0}$ and $\sigma_{\mathrm{F}_{1}}=\sigma_{0}+\Delta\sigma_{T}$
are the deviations. Assuming no prior information about the subject
location, it is also $\Pr\left(\mathrm{F}_{0}\right)=\Pr\left(\mathrm{F}_{1}\right)=1/2$.
Using the log-normal model (\ref{eq:received_power}), (\ref{eq:llr})
can be rewritten as: 
\begin{equation}
\begin{split}\mathrm{\Gamma}(A_{\textrm{S}})\,=\, & \frac{1}{2}\left(\frac{A_{\textrm{S}}-\mu_{\mathrm{F}_{0}}}{\sigma_{\mathrm{F}_{0}}}\right)^{2}\negthinspace-\,\negthinspace\frac{1}{2}\left(\frac{A_{\textrm{S}}-\mu_{\mathrm{F}_{1}}}{\sigma_{\mathrm{F}_{1}}}\right)^{2}\negthinspace-\negthinspace\mathrm{log}\left(\frac{\sigma_{\mathrm{F}_{1}}}{\sigma_{\mathrm{F}_{0}}}\right).\end{split}
\label{eq:gamma_roc}
\end{equation}

The LLR parameters are obtained from the predictions
$A_{\mathrm{S}}^{(p)}(\ell)$ of Sect. \ref{sec:Diffraction-models},
namely $\mu_{\mathrm{F}_{i}}\approx\mu_{\mathrm{F}_{i}}^{(p)}=1/L_{i}\,\sum_{\ell\in\mathcal{L}_{1}}A_{\mathrm{S}}^{(p)}(\ell)$
and $\sigma_{\mathrm{F}_{i}}\approx\sigma_{\mathrm{F}_{i}}^{(p)}=\sqrt{1/L_{i}\,\sum_{\ell\in\mathcal{L}_{i}}\left[A_{\mathrm{S}}^{(p)}(\ell)-\mu_{\mathrm{F}_{i}}^{(p)}\right]^{2}}$,
for hypotheses $\mathrm{F}_{0}$ and $\mathrm{F}_{1}$. The fading
effects \cite{rampa-2017}, \textit{i.e.} $\Delta h_{T}=0$ are also neglected
to highlight the diffraction terms only. 
For comparison, the LLR parameters are also obtained from measurements,
$\mu_{\mathrm{F}_{i}}\approx\mu_{\mathrm{F}_{i}}^{(m)}$ and $\sigma_{\mathrm{F}_{i}}\approx\sigma_{\mathrm{F}_{i}}^{(m)}$,
by replacing $A_{\mathrm{S}}^{(p)}(\ell)$ with $A_{\mathrm{S}}^{(m)}(\ell)$.

\begin{figure}[!t]
\centering \subfloat{\includegraphics[width=0.94\columnwidth]{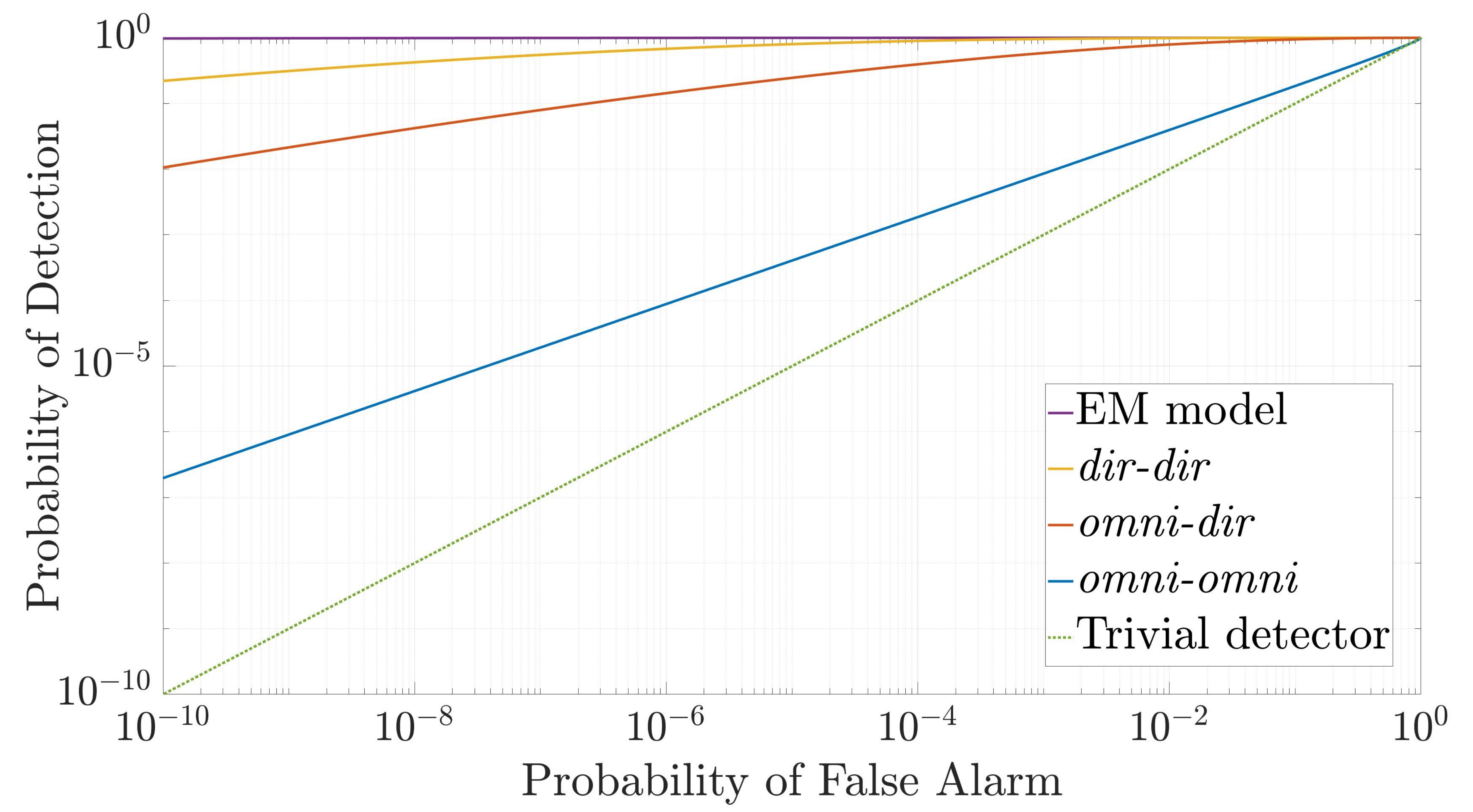}}\caption{ROC plots considering the probabilities related to the EM model, and
to the measurements from the \textit{omni-omni}, \textit{omni-dir}
and \textit{dir-dir} cases. The trivial detector is shown, too.}
\label{fig:roc} \vspace{-0.2cm}
 
\end{figure}

In Fig.~\ref{fig:roc}, we analyze the Receiver Operating Characteristic
(ROC) figures\color{black}\cite{ROC}\color{black}, \color{black}
using the LLR as in (\ref{eq:gamma_roc}), \color{black} for all
scenarios. The ROC associated to the \textit{dir-dir} scenario is
the one with the best performance, being closer to the EM model predictions.
The trivial detector implements a random choice. 
\begin{figure}[!t]
\centering \subfloat{\includegraphics[angle=-90,width=0.9\columnwidth]{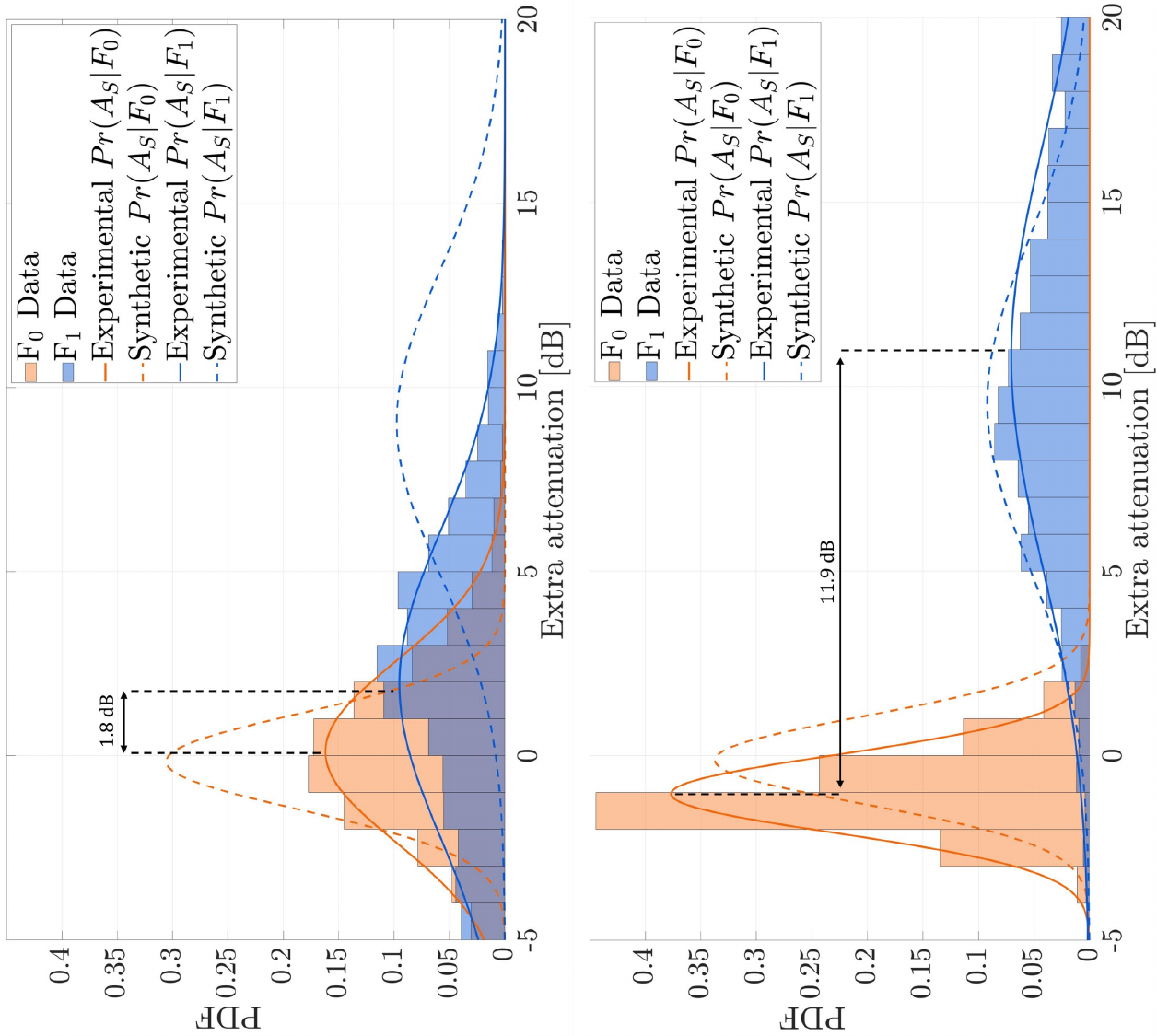}}

\caption{From top to bottom: estimated $\textrm{Pr}(A_{S}|F_{0})$ and $\textrm{Pr}(A_{S}|F_{1})$
from the experimental data and the synthetic (EM) model for the \textit{omni-omni}
(top) and the \textit{dir-dir} scenarios (bottom). Histograms from
experimental data are shown, too.\color{black}}
\label{fig:connect_round_versus_density} 
 \vspace{-0.2cm}
\end{figure}
\begin{table}[tp]
\protect\caption{\label{parameters_oo} \color{black}Likelihood separation and KL divergence.}
\color{black} \vspace{0cm}
\begin{centering}
\begin{tabular}{|c|c|c|}
\cline{2-3} 
\multicolumn{1}{c|}{} & \multicolumn{1}{c|}{\textbf{EM models} ($A_{\mathrm{S}}^{(p)}$)} & \multicolumn{1}{c|}{\textbf{Measurements }($A_{\mathrm{S}}^{(m)}$)}\tabularnewline
\multicolumn{1}{c|}{} & (\textbf{\textit{omni-omni}}$/$\textbf{\textit{dir-dir}})  & (\textbf{\textit{omni-omni}}$/$\textbf{\textit{omni-dir}}$/$\textbf{\textit{dir-dir}})\tabularnewline
\hline 
\multicolumn{1}{|c|}{\textbf{$\mu_{\mathrm{F}_{0}}-\mu_{\mathrm{F}_{1}}$}} & $9.2$~dB$/$$9.6$~dB  & $1.8$~dB$/$$6.7$~dB$/$$11.9$~dB \tabularnewline
\hline 
\multicolumn{1}{|c|}{\textbf{$\mathrm{KL}(\mathrm{F}_{0}||\mathrm{F}_{1})$}} & $2.59$$/$$2.60$  & $1\mathrm{e}{-4}$$/$$0.69$$/$$2.44$ \tabularnewline
\hline 
\end{tabular}
\par\end{centering}
\medskip{}
 \vspace{-0.2cm}
\end{table}
\color{black}Considering that ROC performances depend on the LLR decision regions, \textit{i.e.} the
separation of the log-likelihood (LL) functions \cite{ROC}, in Fig.~\ref{fig:connect_round_versus_density},
we compare the LLs $\Pr\left(A_{\mathrm{S}}\,|\mathrm{F}_{1}\right)$
and $\Pr\left(A_{\mathrm{S}}\,|\mathrm{F}_{0}\right)$ for \textit{omni-omni}
(top) and \textit{dir-dir} scenarios (bottom) obtained from experimental
data ($\mu_{\mathrm{F}_{i}}^{(m)}$,$\sigma_{\mathrm{F}_{i}}^{(m)}$)
and predictions $(\mu_{\mathrm{F}_{i}}^{(p)},\sigma_{\mathrm{F}_{i}}^{(p)})$
using synthetic data, respectively. In Tab.~\ref{parameters_oo}
we also report the average LL separation $\mu_{\mathrm{F}_{0}}-\mu_{\mathrm{F}_{1}}$
and the corresponding Kullback-Leibler (KL) divergence \cite{kull} using measured and
predicted parameters. 
\color{black} The decision regions for the
\textit{dir-dir} scenario are well separated (about $\mu_{\mathrm{F}_{0}}^{(m)}-\mu_{\mathrm{F}_{1}}^{(m)}=11.9$~dB) and this is confirmed by the \emph{dir} model (\ref{eq:V_V0_approx-single})
as $\mu_{\mathrm{F}_{0}}^{(p)}-\mu_{\mathrm{F}_{1}}^{(p)}=9.6$~dB.
\color{black}Similarly, a KL divergence of $2.44$ is predicted against
the measured one of $2.6$. The decision regions for the \emph{omni-omni}
setup are almost overlapped, with average separation of $1.8$~dB
and negligible KL divergence due to the multipath effects and the
absence of any angular filtering. Such effects are not captured by
the \emph{omni} model which performs poorly.
\color{black} 

\section{Conclusions\label{sec:Conclusions}}

This letter proposes a human-body model that accounts for antennas
with non-isotropic radiation characteristics and evaluates the impact
of the radiation pattern for passive radio sensing. Diffraction and
multipath components, that contribute to radio sensing accuracy, are
evaluated experimentally in an indoor environment with mixed antenna
configurations.

The angular filtering properties of directional antennas mitigate
the multipath effects and make the propagation scenario closer to
the results predicted by the diffraction-based EM model. Considering
the problem of classifying target proximity, the model effectively
predicts the separation of the decision regions, observed with directional
antennas, for target inside or outside the Fresnel's ellipsoid. On
the contrary, using omnidirectional antennas, the multipath effects dominate
over diffraction and the model fails to predict such separation. 

\color{black} Future works will adapt the proposed model to Wireless
LAN sensing devices leveraging antennas with software re-configurable
radiation characteristics.

\end{document}